\documentclass[10pt,conference]{IEEEtran}
\IEEEoverridecommandlockouts

\usepackage{xcolor}
\definecolor{improvgreen}{RGB}{34,139,34}
\definecolor{regressred}{RGB}{178,34,34}
\usepackage{tabularray}
\usepackage{xspace}

\usepackage{amsmath,amssymb,amsfonts}
\usepackage{algorithm}
\usepackage{algorithmic}
\usepackage{graphicx}
\usepackage{textcomp}
\usepackage{xcolor}
\usepackage{verbatim}

\usepackage{tikz}
\usetikzlibrary{calc, positioning, arrows.meta, shadows}
\usepackage{makecell}

\DeclareMathOperator*{\argmin}{arg\,min}   

\usepackage[numbers,sort&compress]{natbib}

\newcommand{\gbox}[3]{
  \draw[grp] ([xshift=-2.5mm,yshift=-2.8mm]#1.south west)
             rectangle ([xshift=2.5mm,yshift=2.8mm]#2.north east);
  \ifx\relax#3\relax\else
    \node[glab] at ([yshift=5mm]$(#1.north)!0.5!(#2.north)$) {#3};
  \fi
}

\newcommand{\agentp}{$\mathcal{AP}$\xspace}
\newcommand{\minp}{$\mathcal{MP}$\xspace}
\newcommand{\algo}{\textsc{Trim}\xspace}
\newcommand{\algoG}{\textsc{TRIM-G}\xspace}
\newcommand{\algoNG}{\textsc{TRIM-NG}\xspace}

\newcommand{\mini}{\textsc{MiniSWE Agent}\xspace}
\newcommand{\swe}{\textsc{SWE Agent}\xspace}
\newcommand{\crashfixer}{\textsc{CrashFixer}\xspace}
\newcommand{\openhands}{\textsc{OpenHands}\xspace}
\newcommand{\aislop}{\textsc{CodeSlop}\xspace}

\newcommand{\patchmin}{\textsc{Patch Minimization}\xspace}
\newcommand{\patchdiff}{$\Delta_{Slop}$\xspace}

\newcommand*\blackcircled[1]{\tikz[baseline=(char.base)]{
    \node[shape=circle, fill=black, text=white, inner sep=0.75pt, draw=black] (char) {#1};}}

\usepackage{amsmath,amssymb,mathtools}
\usepackage{amsmath,amssymb,amsfonts}
\usepackage{amsthm}
\newtheorem{definition}{Definition}

\usepackage{booktabs}   
\usepackage{subcaption}
\usepackage{booktabs}
\usepackage[table]{xcolor}
\usepackage{multirow}

\usepackage{hyperref}
\usepackage{cleveref}

\usepackage[most]{tcolorbox}
\newtcolorbox{rqsummary}[1]{
  enhanced,
  colback=gray!8,
  colframe=gray!65,
  coltitle=black,
  colbacktitle=gray!25,
  boxrule=0.7pt,
  arc=1mm,
  left=1.5mm,
  right=1.5mm,
  top=1mm,
  bottom=1mm,
  fonttitle=\bfseries,
  title=\small{#1}
}

\usepackage{pgfplots}
\pgfplotsset{compat=1.18}

\newcommand{\ie}{\textit{i.e.,}\xspace}

\definecolor{commentgray}{gray}{0.45}   
\definecolor{headerblue}{RGB}{225,238,255}

\newcommand{\LineComment}[1]{\hfill{\color{blue}\(\triangleright\)~\textit{#1}}}

\def\BibTeX{{\rm B\kern-.05em{\sc i\kern-.025em b}\kern-.08em
    T\kern-.1667em\lower.7ex\hbox{E}\kern-.125emX}}

\newcommand{\authblk}[2]{%
  \begin{minipage}[t]{0.31\textwidth}\centering
  #1\\[0.4ex] #2%
  \end{minipage}}

\begin{document}

\title{\textsc{TRIM}: Reducing AI-Generated \textsc{CodeSlop} via \\ Agent Trajectory Minimization\\
\thanks{}
}


\author{%
\authblk{Alex Mathai}{\textit{Dept. of Computer Science}\\ \textit{Columbia University},\\ New York City, USA\\ alexmathai@cs.columbia.edu}\hfil
\authblk{Shobini Iyer}{\textit{Dept. of Computer Science}\\ \textit{Columbia University},\\ New York City, USA\\ si2449@columbia.edu}\hfil
\authblk{Aleksandr Nogikh}{\textit{Google Inc}\\ München, Germany\\ nogikh@google.com}\\[3ex]
\authblk{Petros Maniatis}{\textit{Google DeepMind}\\ Mountain View, CA, USA\\ maniatis@google.com}\hfil
\authblk{Franjo Ivan\v{c}i\'{c}}{\textit{Google Inc}\\ Princeton, NJ, USA\\ ivancic@google.com}\hfil
\authblk{Junfeng Yang}{\textit{Dept. of Computer Science}\\ \textit{Columbia University},\\ New York City, USA\\ junfeng@cs.columbia.edu}\\[3ex]
\authblk{Baishakhi Ray}{\textit{Dept. of Computer Science}\\ \textit{Columbia University},\\ New York City, USA\\ rayb@cs.columbia.edu}%
}

\maketitle
\thispagestyle{plain}

\begin{abstract}

Coding agents are increasingly used to accelerate code generation in many downstream tasks, such as fixing bugs, building applications, and prototyping. However, despite their value as coding assistants, agent-generated code tends to be larger and more verbose than the corresponding human-written implementation. In this work, we show that the cause lies in the agent's own search process: while iterating toward a passing solution, an agent accumulates speculative edits, abandoned hypotheses, and temporary changes that persist into the final patch. This may seem harmless for a single patch, but the problem compounds as agents take responsibility for ever-larger portions of a codebase---a codebase that was once minimal and well-maintained slowly accumulates redundancy faster than it can be cleaned up, drifting to a state that is harder to maintain. 

Given the magnitude of this problem, we take a step towards alleviating this issue. First, we formally define this phenomenon as \textsc{CodeSlop}---the residual and functionally unnecessary edits commonly seen in AI-generated code. 
We then introduce our algorithm \textsc{TRIM} (\underline{T}rajectory-guided \underline{R}edundancy \underline{I}dentification and \underline{M}inimization). Rather than minimizing \textsc{CodeSlop} directly, \textsc{TRIM} instead minimizes agent trajectories.
As we show empirically, this indirect technique of minimizing \textsc{CodeSlop} is highly effective: \textsc{TRIM} cuts \textsc{CodeSlop} by 17.9\%--32.9\% across agentic scaffolds, with negligible performance regression. \textsc{Trim} is also highly efficient, requiring roughly half the validation cost of algorithmic baselines such as Delta Debugging.

\end{abstract}


\section{Introduction}
\label{sec:intro}

Modern coding agents are rapidly transforming software development~\cite{claudecode, GithubCopilot, CursorAi}. They can navigate repositories, write code, run tests, and iteratively repair failures, increasingly producing working patches with little human intervention. As these systems become more capable, the challenge is no longer simply generating correct code, but understanding and maintaining it afterward~\cite{roychoudhury2026agentic}. 
Yet most work on coding agents focuses on a single question:
\textit{Does the patch pass the tests?} However, test passing alone says
little about the quality of the resulting patch---developers must still
review, understand, and maintain the generated changes.

This challenge is already emerging in practice. Although many
agent-generated patches successfully pass their tests, developers
increasingly report that many of the generated changes are unnecessary,
requiring additional effort to review, simplify, and sometimes even reject
such patches before they can be
merged~\cite{ceka2025understanding,nakashima2026agentic,asdaque2026novice}.
To understand how these unnecessary changes arise, we investigate program
repair---one of the most mature applications of coding agents---where agents
iteratively edit code and execute tests until a patch eventually succeeds by
passing the required test suite. We refer to this sequence of edits and
validation steps as the agent's \emph{repair trajectory}. The repair
trajectory provides a natural lens for understanding how unnecessary changes
accumulate during the agent's exploration for a successful fix.


When we examined these repair trajectories, one pattern stood out. Agents
typically arrive at a correct repair only after several rounds of editing and
testing---an expected aspect of agentic software
engineering~\cite{kim2026trajeval,xiao2025trajectory}. What surprised us was
that agents often retain, rather than discard, the exploratory changes that
led to a successful repair. As a result, the final patch contains not only
the needed edits that resolve the defect, but also remnants of the search process.


Figure~\ref{fig:kernel_motivating_example} illustrates this phenomenon. When resolving a Linux kernel vulnerability \cite{mem_leak}, \swe \cite{yang2024swe} produces 21 modified lines across three feedback cycles, yet \textit{only three} of those lines are required for the fix. The remaining changes are remnants of the search process---speculative edits, abandoned hypotheses, and temporary modifications that persist simply because the agent has no reason to remove them once a passing solution is found. Interestingly, these three lines happen to coincide with the ground-truth human fix. 
We refer to these residual search artifacts that survive into the final patch as \emph{\aislop}.

\aislop\ not only increases the size of agent-generated patches, but also
buries the changes that actually resolve the defect beneath exploratory
edits, making patches harder to review, understand, and
maintain~\cite{watanabe2026cut,nakashima2026agentic}. More concerning, this
is not an isolated phenomenon. Across our benchmarks, agent-generated
patches routinely contain substantial amounts of \aislop, consistent with
recent reports that AI-generated code is increasingly verbose and
accumulates redundancy faster than human-written
code~\cite{orlanski2026slop,dou2026wrong,he2025speed,ehsani2026faster}.

\textbf{{Defining \aislop.}}
Once this phenomenon is recognized, the next task is to define it.
Existing work characterizes slop through static properties of the resulting
code artifact, such as verbosity, redundancy, or the accumulation of
complexity over time~\cite{orlanski2026slop,dou2026wrong,abbassi2025taxonomy}.
Our key insight is that, for agent patches, what matters is not
how the retained code looks, but whether it is actually necessary for the
repair. A change may be readable, well structured, and free of
duplication, yet still be removable without affecting the final solution. We
therefore define \aislop\ as \emph{removable functional redundancy}:
changes introduced during the agent's search process that can be removed
while preserving the successful repair. 
To our knowledge, this is the first work to formalize \aislop\ through a
functional, rather than static, definition.

\textbf{Identifying \& Reducing \aislop:~Challenges.} 
Defining \aislop\ is only the first step; identifying it is substantially
harder. Once a patch passes its tests, both the essential changes and the
residual search artifacts appear equally valid, making them difficult to
distinguish from the final patch alone. Moreover, dependencies among changes
mean that the necessity of one edit often depends on the presence of others,
preventing edits from being evaluated independently. This challenge is
difficult even for state-of-the-art coding agents. Although agents can
generate successful repairs, they are not trained to identify the minimal set
of changes required for a solution. We confirm this empirically by prompting
agents to minimize their own patches: in $3.8\%$--$44.9\%$ of cases, the
resulting patch either fails to preserve the original behavior or is larger
than the original patch.

\textbf{{Our Solution.}} 
Consequently, rather than searching for \aislop\ over the final patch alone,
we shift the search space to the agent's repair trajectory, where the
temporal ordering of edits provides a natural approximation of their
dependencies. To this end, we formulate \aislop\ identification as a \emph{hierarchical
counterfactual search}: at each level, \algo (\underline{T}rajectory-guided \underline{R}edundancy \underline{I}dentification and \underline{M}inimization) asks whether an entire group of
changes can be removed while preserving the behavior of the successful
repair. The search progressively refines from coarse trajectory groups to
finer-grained changes, validating every candidate removal through execution.
By eliminating large groups of exploratory edits first, \algo\ rapidly
shrinks the search space before reasoning about individual changes.

Such agent trajectory guided exploration differs fundamentally from traditional program
minimization techniques such as
Delta Debugging~\cite{10.1145/318774.318946} and
\texttt{git-bisect}~\cite{git_bisect}. These techniques search across
developer-authored versions or commits, whose histories are relatively clean
and largely reflect intentional software evolution. In contrast, repair
trajectories record an inherently exploratory process in which edits are
proposed, validated, refined, reverted, and superseded before a successful
repair is found. Rather than rediscovering these dependencies through
exhaustive counterfactual testing, \algo\ exploits the trajectory's 
hierarchical structure directly, producing a minimal patch that preserves the
successful repair while removing artifacts of the agent's search process.


\textbf{\textit{Results.}} We evaluate our approach across four agent scaffolds (\crashfixer, \swe, \mini, and \openhands) on Live-kBench \cite{huang2026outrunningllmcutoffslive} and \textsc{Swe-Bench}\cite{jimenez2023swe}. Our approach reduces \aislop by $17.8\%$--$32.9\%$, achieving a $1.6$$\times$--$3.1$$\times$ improvement over agent-based minimization baselines while introducing negligible regressions in correctness. 
Minimization even brings some agent-generated patches into exact agreement with the developer-written patch.

In summary, this paper makes the following contributions:

\begin{itemize}
\item We introduce \aislop, a new formulation of removable functional redundancy in agent-generated patches.

\item We develop \algo, a trajectory-aware algorithm for reducing \aislop.

\item We empirically show that substantial amounts of \aislop can be removed from agent-generated patches across multiple agents and benchmarks.

\end{itemize}

\section{Motivating Example} 
\label{sec:motiv}

\begin{figure}[t]
\centering

\begin{subfigure}{0.95\linewidth}
  \centering
  \includegraphics[width=\linewidth]{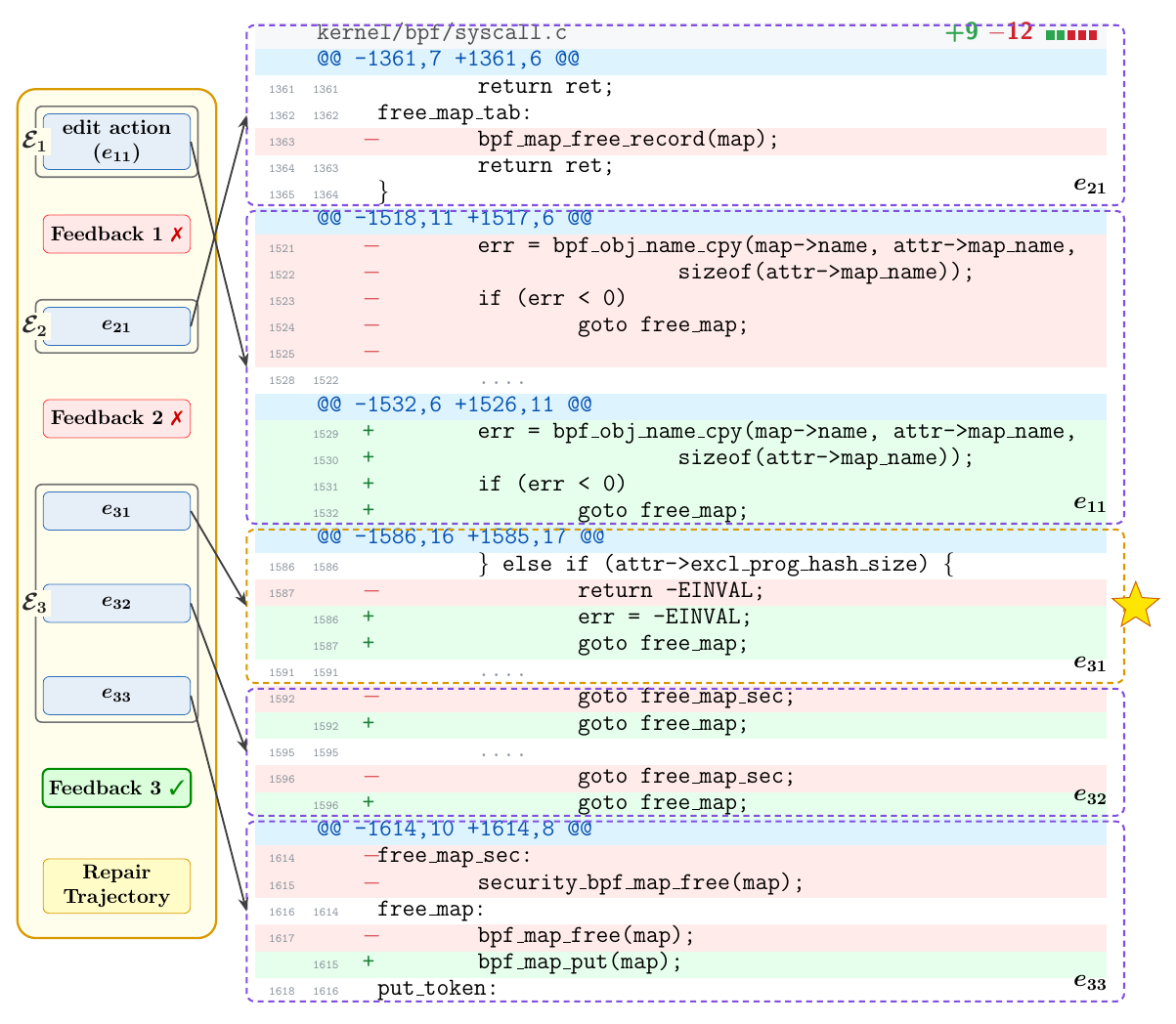}
  \caption{Linux vulnerability~\cite{mem_leak}; here the human patch is edit $e_{31}$.}
  \label{fig:kernel_motivating_example}
\end{subfigure}

\vspace{1.5mm}

\begin{subfigure}{0.95\linewidth}
  \centering
  \includegraphics[width=\linewidth]{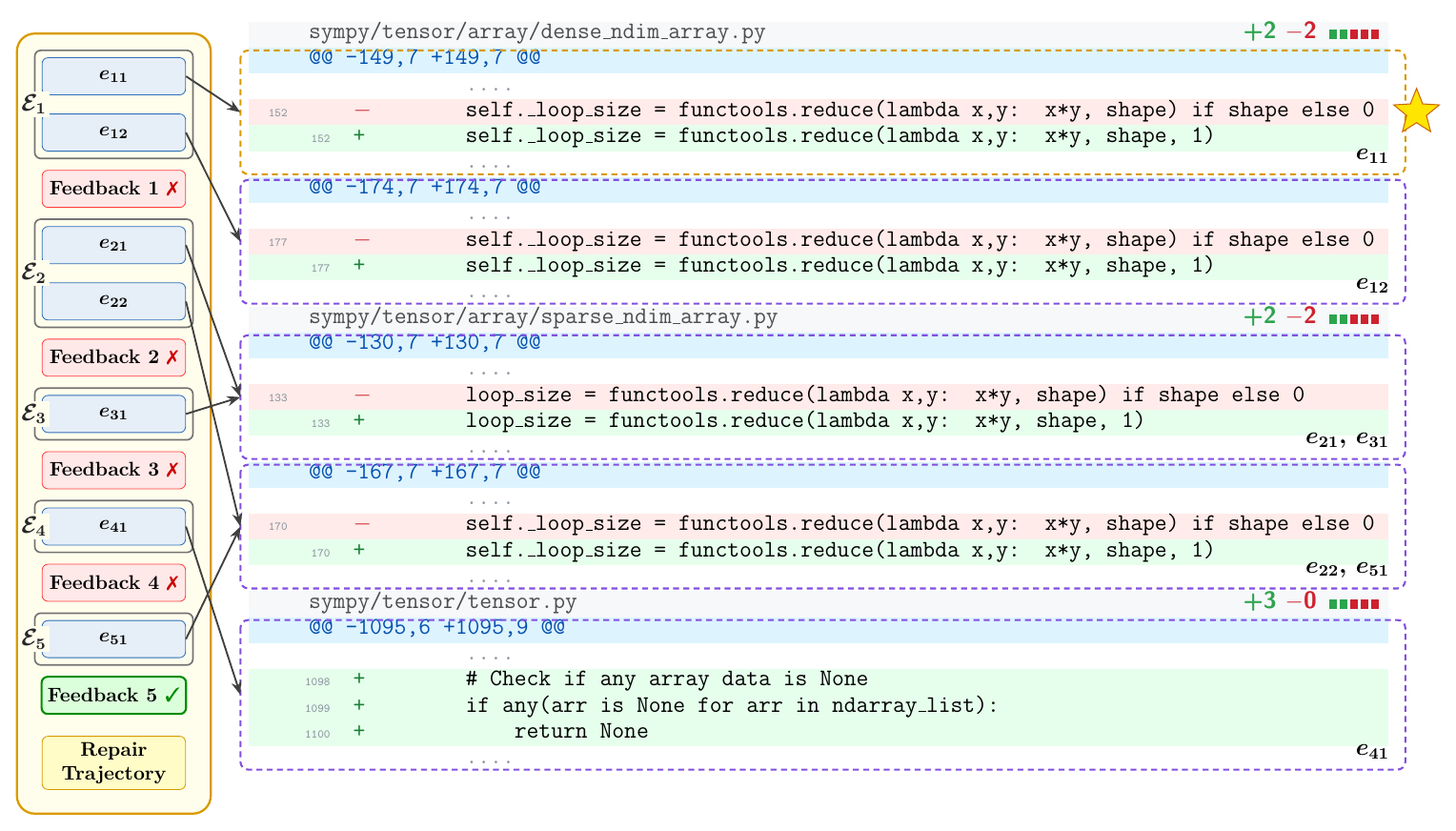}
  \caption{SWE-Bench \texttt{sympy\_\_sympy-15017}.
  The agent patch spans three files and five hunks, yet the true fix is a single
  line ($e_{11}$); \algo eliminates two files and prunes the patch down to that one edit,
  identical to the human fix.}
  \label{fig:sympy_motivating_example}

\vspace{-1mm}
\end{subfigure}

\caption{\small{\textbf{Two motivating examples of \aislop.} 
\algo recovers the human-equivalent fix hidden inside a
larger agent patch. Each trajectory shows (i) sequence of edits ($\mathcal{E}$), 
(ii) atomic edit action ($e_{ij}$), and (iii) Feedbacks. 
Atomic  edits in the trajectory map to regions of the final patch (bounded boxes).
The starred region ($\star$) is the human patch, while the purple-bordered regions are
\aislop---residual edits ($e_{ij}$) from the agent's search process that \algo prunes away.}}
\label{fig:motivating_examples}
\vspace{-4mm}
\end{figure}

In this section, we use a real-world repair trajectory to illustrate how
\aislop\ emerges during agent-driven program repair.
Figure~\ref{fig:kernel_motivating_example} shows the repair trajectory and
final patch generated by \swe~\cite{yang2024swe} while fixing a Linux kernel
memory leak vulnerability~\cite{mem_leak}.

Now consider a developer reviewing the resulting patch
(Figure~\ref{fig:kernel_motivating_example}, right). Although the patch
correctly fixes the vulnerability, it modifies a single file through $21$
changed lines spanning five hunks. Every one of these changes demands
attention: the developer must understand what it does, why it was introduced,
and whether it interacts with the rest of the patch. Consequently,
unnecessary edits significantly increase the cognitive effort required to
review and maintain the repair.

To understand where these unnecessary changes originate from, we examine the
repair trajectory (Figure~\ref{fig:kernel_motivating_example}, left). Rather
than discovering the fix immediately, the agent progresses through three edit
sequences ($\mathcal{E}_{1}$--$\mathcal{E}_{3}$), each followed by test
execution. The first two fail validation, while the third finally produces a
passing repair. Figure~\ref{fig:kernel_motivating_example} maps these edit
sequences to the final cumulative patch, showing how changes introduced
during earlier exploration persist in the final solution.

At first glance, every modification appears equally important because all are
present in the passing patch. A closer inspection, however, reveals that only
the starred edit ($e_{31}$) is actually required to resolve the vulnerability
and matches the developer-written fix. The remaining edits neither contribute
to the repair nor affect correctness; they simply persist because the agent
has no incentive to remove them once a passing solution is found. We refer to
these residual search artifacts as \aislop.

\textbf{Trajectory-Guided Search.}
The repair trajectory provides the missing information needed to distinguish
essential changes from \aislop. Rather than blindly searching over the
$2^5=32$ possible subsets of the five patch hunks, \algo\ shifts the search
space to the repair trajectory and performs a hierarchical counterfactual
search. It first reasons at the granularity of edit sequences, asking whether
each edit sequence is necessary to preserve the successful repair. In this
example, it first removes $\mathcal{E}_1$ and validates that the repair still
succeeds, then repeats the process for $\mathcal{E}_2$. Since neither
sequence affects the outcome, both are discarded. The search then descends
into the remaining edit sequence, $\mathcal{E}_3$, and repeats the same
counterfactual reasoning over its edit actions, successively removing
$e_{33}$ and $e_{32}$ until only the essential edit $e_{31}$ remains.
In reality, the ordering of the sequences determine how fast \algo will converge.

This search fundamentally differs from traditional techniques such as Delta
Debugging~\cite{10.1145/318774.318946}. Without access to the repair
trajectory, Delta Debugging treats the final patch as an unstructured
collection of hunks and must recover edit dependencies by exploring
combinations of hunk subsets. Consequently, it operates over a substantially
larger search space and converges more slowly, requiring many more
counterfactual validations. In contrast, repair trajectories naturally group
dependent edits according to the agent's exploratory search process. By
eliminating entire trajectory groups before refining to individual edits,
\algo\ prunes large portions of the search space and reaches the same minimal
repair with substantially fewer validation calls.

Figure~\ref{fig:sympy_motivating_example} illustrates another instance of \aislop\ from SWE-Bench.

\section{Problem Formulation}
\label{sec:problem_formulation}

This section formalizes the notion of \aislop and the corresponding
optimization problem.

\subsection{Minimal Behavior-Preserving Patch}
\label{sobsec:minimal_patch}

As illustrated in~\Cref{sec:motiv}, agent-generated patches often contain
modifications accumulated during the agent's search process for a correct
solution, many of which are ultimately unnecessary for satisfying the task.
Our objective is therefore not merely to produce a passing patch, but to
identify the \emph{smallest} patch that still correctly satisfies the task.

A \emph{patch} is a collection of source-code modifications (i.e., added and
deleted lines) that transforms one program version into another. Let
$\mathcal{AP}$ denote an agent-generated patch for a task $T$. We define
$\operatorname{len}(\mathcal{P})$ as the length of a patch $\mathcal{P}$,
measured by the total number of modified lines (added and deleted). Throughout
this paper, the \emph{behavior} of a patch refers to its externally observable
effects relevant to task $T$, such as producing the intended outputs and
satisfying the required correctness conditions, while abstracting away
incidental implementation details.

\begin{definition}[Minimal Behavior-Preserving Patch]

Let $\mathcal{D}(\mathcal{AP},T)$ denote the set of all patches obtained by
removing one or more modifications from $\mathcal{AP}$ while still correctly
satisfying task $T$. The \emph{minimal behavior-preserving patch}, denoted by
$\mathcal{AP}^{*}$, is defined as
\[
\mathcal{AP}^{*}
=
\argmin_{\mathcal{MP}\in\mathcal{D}(\mathcal{AP},T)}
\operatorname{len}(\mathcal{MP}).
\]

That is, $\mathcal{AP}^{*}$ is the shortest patch derivable from
$\mathcal{AP}$ that still correctly satisfies task $T$.

\end{definition}

Intuitively, $\mathcal{AP}^*$ represents the smallest correct realization of the
agent-generated solution for the task $T$ that contains only the modifications necessary for
implementing the task, while removing functionally unnecessary modifications
introduced during the agent's exploration process~\footnote{This formulation deliberately restricts the search space to patches derived from the agent-generated patch. Rather than synthesizing a new solution from scratch, minimization asks which parts of the agent's existing solution are unnecessary. This makes the optimization tractable: the algorithm searches over \agentp instead of the much larger space of all patches that could solve $T$.}. As such, $\mathcal{AP}^*$ defines the \emph{theoretical minimum} patch
under our formulation and serves as the ideal target against which practical
minimization algorithms are measured.

The notion of ``preserving the intended functionality'' is an ideal
specification independent of any particular implementation or evaluation
methodology. In practice, since the true functionality is unobservable,
practical minimization algorithms approximate this ideal by validating
candidate patches using the available execution environment and test suite
(\S\ref{sec:method}).


\subsection{\aislop}

\begin{definition}[\aislop]
Given an agent-generated patch \agentp and its corresponding minimal
behavior-preserving patch $\mathcal{AP}^*$, the amount of \aislop\ in \agentp is
\begin{equation}
    \textsc{CodeSlop}\xspace(\mathcal{AP}) = \operatorname{len}(\mathcal{AP}) - \operatorname{len}(\mathcal{AP}^*)
\end{equation}
Equivalently, $\textsc{CodeSlop}\xspace(\mathcal{AP})$ measures the amount of modifications in \agentp that can be
removed without changing the behavior of the original agent patch with respect to task $T$.
\end{definition}


Unlike traditional notions of code quality, which define slop using static
properties of the final patch (e.g., verbosity, duplication, or structural
complexity~\cite{orlanski2026slopcodebench}), \aislop\ is defined
behaviorally: modifications that can be removed without changing the
behavior of the original agent-generated patch.

\subsection{Problem Statement}
\label{subsec:problem_statement}

The preceding definitions characterize the ideal notions of \aislop\ and its
corresponding minimal behavior-preserving patch $\mathcal{AP}^*$. Based on them, we now formulate the
computational problem addressed in this paper.

\begin{definition}[\textsc{Patch Minimization}]
Given an agent-generated patch \agentp, \textsc{Patch Minimization} seeks to
recover its corresponding minimal behavior-preserving patch  $\mathcal{AP}^*$.
\end{definition}

Recovering $\mathcal{AP}^*$ requires identifying the smallest behavior-preserving patch among all candidate patches derivable from \agentp. Each
modification in \agentp may either be retained or removed,
yielding an exponential search space of size $2^n$, where $n$ is the number of
modifiable units in the patch. Consequently, exhaustive search is intractable when n is large.

Moreover, the ideal notion of behavior preservation cannot be established
directly. Instead, coding agents validate candidate patches by executing them
in an execution environment ($Env$) using a set of executable test files. Consequently, practical algorithms compute \minp---an \emph{approximation}
 of $\mathcal{AP}^*$ by searching for a minimized patch satisfying:

\begin{enumerate}
    \item \minp exhibits the same observable behavior as \agentp under $Env$ when
          executing test cases; 
    \item $\operatorname{len}(\mathcal{MP}) \leq \operatorname{len}(\mathcal{AP})$.
\end{enumerate}

\begin{figure}[t]
    \centering
    \includegraphics[width=\columnwidth]{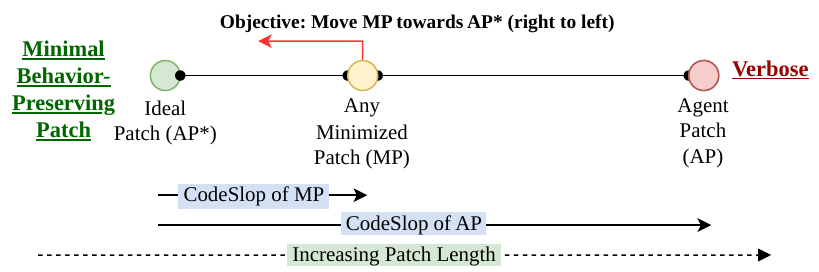}
    \caption{ The task of minimizing \aislop. The horizontal line measures patch length. On the left is the minimal behavior-preserving patch ($\mathcal{AP}^*$) and on the right is the original verbose agent patch (\agentp). In the middle is the minimized patch (\minp). To minimize \aislop and patch length, we move \minp closer to $\mathcal{AP}^*$ and further away from \agentp.
    }
    \label{fig:problem_formulation}
    \vspace{-10pt}
\end{figure}


Since the ideal optimum $\mathcal{AP}^{*}$ is generally unknown, practical
algorithms instead compute an approximation $\mathcal{MP}$. The quality of an approximation is determined by its proximity to $\mathcal{AP}^*$: the
closer \minp is to $\mathcal{AP}^*$, the more \aislop\ has been eliminated
(see~\Cref{fig:problem_formulation}).

We therefore measure the amount of \aislop\ eliminated by a minimization procedure as
$\Delta_{\textsc{Slop}}
=
\operatorname{len}(\mathcal{AP})
-
\operatorname{len}(\mathcal{MP})$
which corresponds to the reduction in patch length achieved by the
minimization algorithm (see~\Cref{fig:problem_formulation}). 

\textbf{Note.} The repair trajectory is not part of the definition of
\aislop. The definition depends only on behavioral equivalence. The
trajectory instead provides structural information that enables efficient
identification of behaviorally redundant changes, as we will discuss next.

\section{Methodology}
\label{sec:method}

This section presents \algo, an approximation algorithm for Patch Minimization. We start with a high-level overview (\S\ref{subsec:system_overview}) of the algorithm. We then elaborate on each of its components in the rest of the section.

\begin{figure*}[t]
  \centering
  \includegraphics[width=0.9\textwidth]{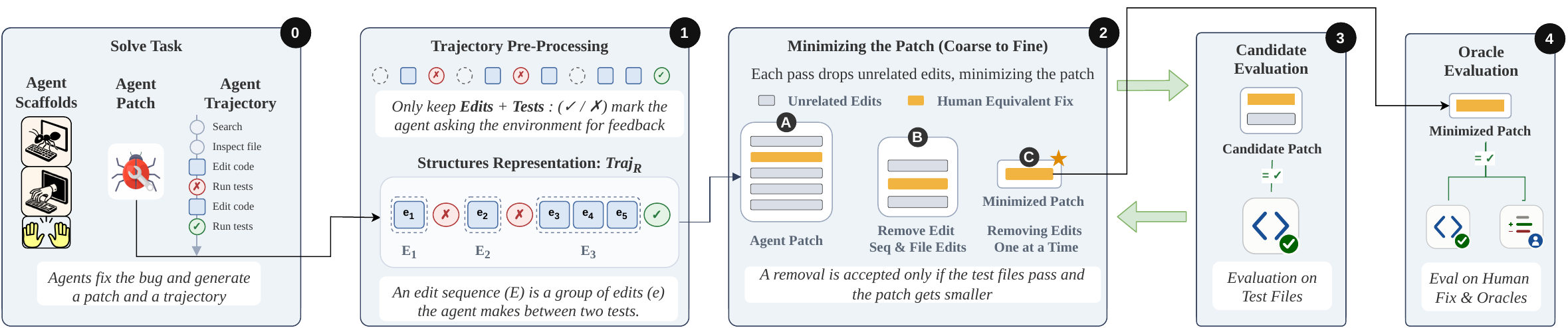}
  \caption{
  \small{Overview of \algo.
~\protect\blackcircled{0} A coding agent repairs a bug, producing a patch and an execution trajectory containing code edits, repository interactions, and feedback requests ($FR$).
~\protect\blackcircled{1} We preprocess the trajectory, retaining only edits and feedback requests, and group edits between consecutive $FR$s into \emph{edit sequences} ($\mathcal{E}_1$--$\mathcal{E}_3$), forming the trajectory-derived search space $Traj_R$ (shown for the motivating example in \S\ref{sec:motiv}).
~\protect\blackcircled{2} \algo\ performs hierarchical counterfactual search, progressively removing edit sequences, files, and finally individual edits.
~\protect\blackcircled{3} A removal is accepted only if the resulting patch passes validation while reducing patch size. The orange line denotes the essential fix; gray lines represent \aislop\ eliminated during minimization.
~\protect\blackcircled{4} The minimized patch is evaluated against hidden oracle tests and, optionally, compared with the developer-written patch.}
  }
  \label{fig:system_overview}
\vspace{-5mm}
\end{figure*}

\subsection{System Overview}
\label{subsec:system_overview}

As formulated in Section~\ref{sec:problem_formulation}, \patchmin\ seeks the
minimal behavior-preserving patch $\mathcal{AP}^*$. We formulate this as a
\emph{counterfactual reasoning} problem: determining whether a modification
is necessary requires asking, \emph{what would happen if it had never been
made?} Each counterfactual corresponds to removing one or more modifications
from \agentp and validating whether the resulting
patch preserves the desired behavior. The challenge is that these
modifications are highly dependent---the effect of removing one
modification often depends on the presence of others. Consequently, solving
\patchmin\ exactly requires searching an exponential space of candidate
patches ($2^n$ assuming $n$ modifications), making exhaustive search
computationally intractable.

Our key insight is that the agent's repair trajectory provides the
structure needed to explore these dependent counterfactuals efficiently.
Rather than searching arbitrary subsets of the final patch, \algo\
(\textbf{T}rajectory-guided \textbf{R}edundancy \textbf{I}dentification
and \textbf{M}inimization) organizes modifications according to the
trajectory's natural hierarchy and progressively eliminates groups of
related changes while preserving the original behavior.
At a high level, \algo\ consists of three stages
(Figure~\ref{fig:system_overview}):

\noindent\protect\blackcircled{1}~\textit{Trajectory-guided search space construction.}
Starting from the repair trajectory, \algo\ reconstructs the edits that
contribute to the final patch and organizes them into a hierarchical search
space. \protect\blackcircled{2}~\textit{Hierarchical counterfactual search \& validation.}
\algo\ progressively minimizes the patch, reasoning from coarse groups of
related modifications to individual edits. \protect\blackcircled{3} Candidate counterfactual patches
are validated through execution and accepted only if they preserve the
original behavior while reducing patch size.

The remainder of this section describes each stage in detail.

\subsection{Trajectory-Guided Search Space Construction}
\label{subsec:search_space}

The first stage of \algo\ transforms the raw execution trajectory into a
hierarchical search space for counterfactual reasoning. An
\emph{execution trajectory}, $Traj$, records the sequence of interactions
between a coding agent and its execution environment, including code edits,
test executions, repository exploration, and command execution. However,
many of these actions do not contribute to the final agent-generated patch,
while the remaining edits must be organized to expose their dependencies.
This stage therefore extracts the trajectory actions that contribute to the
final patch and organizes them into a hierarchical search space for Patch
Minimization.

\subsubsection{Trajectory Reconstruction}

Not all actions in $Traj$ are relevant for Patch Minimization.
Therefore, \algo first projects the execution trajectory onto a
reduced trajectory, denoted by $Traj_R$ (See~\Cref{fig:kernel_motivating_example}), by retaining only

\begin{itemize}
    \item $\mathcal{E}$: sequences of code edits performed between consecutive
    feedback requests, and
    \item $\mathcal{FR}$: feedback requests corresponding to executions of the
    task-related test suite.
\end{itemize}

All remaining actions, including repository search, file inspection,
directory navigation, and other non-modifying operations, are discarded because
they do not affect the resulting patch. The reduced trajectory is represented
as
\[
Traj_R
=
\left\langle
(\mathcal{E}_1,\mathcal{FR}_1),
(\mathcal{E}_2,\mathcal{FR}_2),
\ldots,
(\mathcal{E}_k,\mathcal{FR}_k)
\right\rangle,
\]

where
\[
\mathcal{E}_i=\langle e_{i1},e_{i2},\ldots,e_{in}\rangle
\]
denotes the {ordered sequence} of \emph{atomic edit actions} ($e_{ij}$) performed after
$\mathcal{FR}_{i-1}$ and before $\mathcal{FR}_i$, where
$\mathcal{FR}_0=\emptyset$. Each atomic edit action $e_{ij}$ corresponds to a single
editing operation issued by the agent (e.g., a search-and-replace operation or
a code insertion/deletion). 
$e_{ij}$ may modify multiple lines or multiple locations within one file using a standard search-and-replace operation. The corresponding $\mathcal{FR}_i$ denotes the
feedback request issued after completing the edit sequence
$\mathcal{E}_i$.

For illustration, consider the reduced trajectory corresponding to the motivating example in~\Cref{fig:kernel_motivating_example}. The reduced trajectory is
$Traj_R=\langle(\mathcal{E}_1,\mathcal{FR}_1),(\mathcal{E}_2,\mathcal{FR}_2),(\mathcal{E}_3,\mathcal{FR}_3)\rangle)$,
where $(\mathcal{E}_1=\langle e_{11}\rangle)$, $(\mathcal{E}_2=\langle e_{21}\rangle)$, and $(\mathcal{E}_3=\langle e_{31},e_{32},e_{33}\rangle)$. As described next, \algo\ performs Patch Minimization by searching over this trajectory representation, progressively refining the search from edit sequences to individual files and, ultimately, individual edit actions.

\textit{Implementation Challenges.}
Coding agents expose heterogeneous editing primitives, ranging from shell
commands (e.g., \texttt{sed}) to structured editing tools (e.g.,
\texttt{replace}). To support a unified minimization algorithm across
different agent scaffolds, \algo\ normalizes every edit into a common
representation consisting of the target file together with its before- and
after-text.

Execution trajectories also record the agent's entire exploratory process,
including edits that are later reverted, overwritten, or otherwise discarded.
To recover the edits that actually contribute to the final patch, \algo\
faithfully replays the trajectory in temporal order, applying both edit and
undo operations. The resulting reduced trajectory $Traj_R$ contains only the
edits that survive into the final patch while preserving their original
ordering.

\subsubsection{Trajectory-derived Search Space}

The reduced trajectory $Traj_R$ defines the search space explored by
\algo. Agents naturally organize repairs into successive
\emph{edit--feedback request} iterations: before issuing a feedback request,
the agent performs a sequence of edit actions corresponding to a single
repair attempt. Accordingly, each pair $(\mathcal{E}_i,\mathcal{FR}_i)$
represents one repair attempt together with its validation. Rather than
searching arbitrary subsets of edits, \algo\ first reasons over these
edit-sequence units before progressively refining the search to individual
files and, ultimately, atomic edit actions.

\subsection{Hierarchical Counterfactual Search}
\label{subsec:method-B}

This section presents the design principles underlying \algo. At a high
level, \algo formulates Patch Minimization as a constrained optimization
problem: remove as much \aislop\ as possible while preserving program
behavior and keeping the minimization process computationally practical.

\subsubsection*{Optimization Objective}

As established in Section~\ref{sec:problem_formulation}, the objective of \algo is to approximate the minimal behavior-preserving patch of an agent-generated patch \agentp\ for a given task $T$. Rather than minimizing \agentp\ as an unstructured collection of edits, \algo performs \emph{trajectory-guided optimization}, exploiting the temporal organization and repair hypotheses encoded in the repair trajectory to identify functionally unnecessary edits. Thus, instead of synthesizing a new repair, \algo seeks a \emph{one-minimal} edit action subset of the original trajectory edits that still satisfies the task (\S\ref{subsec:one_minimal}).

\subsubsection*{Correctness Constraint}

Since behavior preservation cannot be established directly, every accepted
counterfactual patch must continue to pass the task-specific test suite
($TF$), preserving the observable behavior of the original repair.

\subsubsection*{Efficiency Constraint}

Each candidate counterfactual requires re-executing $TF$, making validation
the dominant cost of minimization (e.g., $\sim$30 minutes per Linux kernel
validation~\cite{mathai2024kgym}). Consequently, effectiveness depends not
only on the amount of \aislop\ removed (\patchdiff), but also on the number
of validation executions, motivating our hierarchical coarse-to-fine search.

\begin{algorithm}[t]
  \caption{\textbf{\emph{\algo}}: hierarchical, coarse-to-fine patch minimization}
  \label{alg:patchmin}
  \begin{algorithmic}[1]
    \REQUIRE $Traj_{R} = \langle (\mathcal{E}_1, \mathcal{FR}_1), \ldots, (\mathcal{E}_n, \mathcal{FR}_n) \rangle$, a list of edit sequences with their feedback requests, where each $\mathcal{E}_i = \langle e_{i1}, \ldots, e_{ij} \rangle$; \\ 
      Flag $oneMin$ (enforce 1-minimality guarantee); \textsc{true} $\Rightarrow$ \algo-G, \textsc{false} $\Rightarrow$ \algo-NG 
    \ENSURE minimized patch $\mathcal{MP}$ with $\operatorname{len}(\mathcal{MP}) \le \operatorname{len}(\mathcal{AP})$
    \STATE $S \gets Traj_{R}$ \LineComment{working set of surviving $\langle\mathcal{E},\mathcal{FR}\rangle$ pairs}
    \STATE $\mathit{TF} \gets \{\mathcal{FR}_1, \ldots, \mathcal{FR}_n\}$ \LineComment{validation oracle}
    \FOR{$g \in \langle\, \textsc{edit\ seq},\ \textsc{file},\ \textsc{edit\ action}\,\rangle$} \label{alg:outerloop}
    \STATE \LineComment{coarse $\to$ fine} 
       \STATE {\color{blue}{\(\triangleright\)~\emph{\textsc{edit\ seq}: edits of one edit sequence $\mathcal{E}_i$;\ \textsc{file}: edits of one file (pooled across sequences);\ \textsc{edit\ action}: a single edit action $e_{ij}$}}} \label{alg:group}
      \REPEAT  \label{alg:repeat}
        \STATE \textbf{pool} the surviving edits of $S$ and \textbf{group} them into units $\langle u_1, \ldots, u_m \rangle$ by $g$ \label{alg:regroup}
        \STATE $\mathit{changed} \gets \textsc{false}$
        \FOR{$j \gets m$ \textbf{downto} $1$} 
        \STATE \LineComment{reverse pass}
         \IF{$\operatorname{apply}(S \setminus u_j)$ passes $\mathit{TF}$ \textbf{and} $\operatorname{len}(S \setminus u_j) < \operatorname{len}(S)$} \label{alg:accept}
            \STATE $S \gets S \setminus u_j$; 
            \STATE $\mathit{changed} \gets \textsc{true}$; \LineComment{accept removal} 
         \ENDIF
        \ENDFOR
      \UNTIL{$\lnot\,\mathit{changed} \ \lor\ \lnot\,\mathit{oneMin}$} \LineComment{fixpoint or one pass} \label{alg:termination}
    \ENDFOR
    \STATE $\mathcal{MP} \gets \operatorname{apply}(S)$
    \RETURN $\mathcal{MP}$
  \end{algorithmic}
\end{algorithm}

\begin{figure}[t]
\centering
\resizebox{0.9\columnwidth}{!}{%
\begin{tikzpicture}[
  font=\small,
  e/.style={draw=black!35, line width=0.5pt, rounded corners=2.5pt,
            minimum width=11mm, minimum height=8.5mm, inner sep=2pt, font=\normalsize},
  pill/.style={rounded corners=3pt, inner xsep=4pt, inner ysep=2pt,
            font=\small\bfseries, text=white},
  gdesc/.style={anchor=west, font=\footnotesize, text=black!55},
  glab/.style={font=\small\itshape, text=black!60},
  rowL/.style={anchor=west, align=left, text=black!80},
  sh/.style={drop shadow={shadow xshift=0.4pt, shadow yshift=-0.5pt,
            opacity=0.22, fill=black!60}},
  f1/.style={e, sh, fill=orange!18, draw=orange!60, text=black!85},
  f2/.style={e, sh, fill=blue!12,  draw=blue!55,   text=black!85},
  fix/.style={e, sh, fill=blue!25, draw=blue!75!black, line width=1pt, text=black!90},
  dead/.style={e, fill=black!6, draw=black!30,
            dash pattern=on 1.5pt off 1pt, text=black!50},
  glab/.style={font=\small\itshape, text=black!60},
  grp/.style={rounded corners=4pt, draw=black!40, line width=0.6pt, fill=black!5},
  gdesc/.style={anchor=west, font=\small, text=black!55},
  tip/.style={-{Latex[length=2mm]}, draw=black!45, line width=0.7pt},
]

\path (3.4,0)    node[e, draw=none, fill=none] (a11) {}
      ([xshift=12mm]a11.center) node[e, draw=none, fill=none] (a21) {};

\node[rowL] at (-1.95,0)
  {\normalsize\bfseries Agent patch \agentp\\[2pt]\textcolor{black!50}{\small 7 edits}};

\node[dead] (a11) at (3.4,0) {$e_{11}$};
\node[f1, right=7mm of a11]   (a21) {$e_{21}$};
\node[f2, right=0.8mm of a21] (a22) {$e_{22}$};
\node[dead, right=7mm of a22] (a31) {$e_{31}$};
\node[fix, right=7mm of a31]  (a41) {$e_{41}^{\star}$};
\node[f1, right=0.8mm of a41] (a42) {$e_{42}$};
\node[fix, right=0.8mm of a42] (a43) {$e_{43}^{\star}$};
\gbox{a11}{a11}{$\mathcal{E}_1$}
\gbox{a21}{a22}{$\mathcal{E}_2$}
\gbox{a31}{a31}{$\mathcal{E}_3$}
\gbox{a41}{a43}{$\mathcal{E}_4$}
\node[dead] at (a11) {$e_{11}$};
\node[f1]  at (a21) {$e_{21}$};
\node[f2]  at (a22) {$e_{22}$};
\node[dead] at (a31) {$e_{31}$};
\node[fix] at (a41) {$e_{41}^{\star}$};
\node[f1]  at (a42) {$e_{42}$};
\node[fix] at (a43) {$e_{43}^{\star}$};

\node[rowL] at (-1.95,-2.4)
  {\small\bfseries After Edit Seq removal\\[2pt]\textcolor{black!50}{\footnotesize 5 edits}};
\node[f1] (b21) at (3.4,-2.4) {$e_{21}$};
\node[f1, right=0.8mm of b21] (b42) {$e_{42}$};
\node[f2, right=6mm of b42]   (b22) {$e_{22}$};
\node[fix, right=0.8mm of b22] (b41) {$e_{41}^{\star}$};
\node[fix, right=0.8mm of b41] (b43) {$e_{43}^{\star}$};
\gbox{b21}{b42}{File $f_1$}
\gbox{b22}{b43}{File $f_2$}
\node[f1]  at (b21) {$e_{21}$};
\node[f1]  at (b42) {$e_{42}$};
\node[f2]  at (b22) {$e_{22}$};
\node[fix] at (b41) {$e_{41}^{\star}$};
\node[fix] at (b43) {$e_{43}^{\star}$};

\node[rowL] at (-1.95,-4.8)
  {\small\bfseries After File removal\\[2pt]\textcolor{black!50}{\footnotesize 3 edits}};
\node[f2] (c22) at (3.4,-4.8) {$e_{22}$};
\node[fix, right=7mm of c22] (c41) {$e_{41}^{\star}$};
\node[fix, right=7mm of c41] (c43) {$e_{43}^{\star}$};
\gbox{c22}{c22}{}
\gbox{c41}{c41}{}
\gbox{c43}{c43}{}
\node[f2]  at (c22) {$e_{22}$};
\node[fix] at (c41) {$e_{41}^{\star}$};
\node[fix] at (c43) {$e_{43}^{\star}$};

\node[rowL] at (-1.95,-7.2)
  {\small\bfseries After edit action removal \\ (Minimal patch)\\[2pt]\textcolor{black!50}{\footnotesize 2 edits = fix}};
\node[fix] (d41) at (3.4,-7.2) {$e_{41}^{\star}$};
\node[fix, right=0.8mm of d41] (d43) {$e_{43}^{\star}$};
\draw[grp, draw=blue!60, fill=blue!6]
  ([xshift=-2.5mm,yshift=-2.8mm]d41.south west)
  rectangle ([xshift=2.5mm,yshift=2.8mm]d43.north east);
\node[fix] at (d41) {$e_{41}^{\star}$};
\node[fix] at (d43) {$e_{43}^{\star}$};

\draw[tip] (5.0,-0.7) -- (5.0,-1.7);
\node[pill, fill=purple!65, anchor=west] (p0) at (5.5,-0.95) {Seq};
\node[gdesc] at (p0.east) {\,drop $\mathcal{E}_1,\mathcal{E}_3$};

\draw[tip] (5.0,-3.1) -- (5.0,-4.1);
\node[pill, fill=teal!60!black, anchor=west] (p1) at (5.5,-3.6) {File};
\node[gdesc] at (p1.east) {\,drop $f_1$ ($e_{21},e_{42}$)};

\draw[tip] (5.0,-5.5) -- (5.0,-6.5);
\node[pill, fill=orange!75!black, anchor=west] (p2) at (5.5,-6.0) {Edit};
\node[gdesc] at (p2.east) {\,drop $e_{22}$};

\node[f1,  minimum width=6mm, minimum height=5mm] (l1) at (-1.6,-8.7) {};
\node[gdesc] at (l1.east) {\,$file_1$};
\node[f2,  minimum width=6mm, minimum height=5mm] (l2) at (0.2,-8.7) {};
\node[gdesc] at (l2.east) {\,$file_2$};
\node[fix, minimum width=6mm, minimum height=5mm] (l3) at (2.1,-8.7) {};
\node[gdesc] at (l3.east) {\,fix ($\star$)};
\node[dead,minimum width=6mm, minimum height=5mm] (l4) at (4.4,-8.7) {};
\node[gdesc] at (l4.east) {\,dead};

\end{tikzpicture}%
}
\caption{\small{\algo\ recovering the minimal patch from a 7-edit agent patch.
Different rows represent the granularity \algo\ reasons about at each level---edit sequence, then files,
then individual edit actions. Node color denotes the edited file; starred edits
($e_{41},e_{43}$) are the true edit actions and all others are \aislop. The patch shrinks
$7\!\to\!5\!\to\!3\!\to\!2$.}}
\label{fig:trim-example}
\vspace{-5mm}
\end{figure}
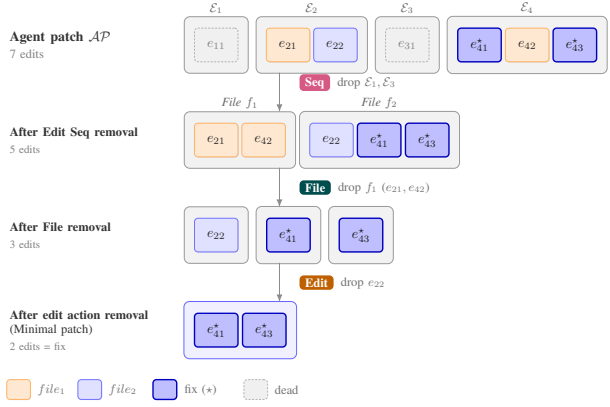

\subsubsection*{Minimization Strategy}

Algorithm~\ref{alg:patchmin} presents the overall hierarchical
counterfactual search procedure.~\Cref{fig:trim-example} also pictorially describes how the algorithm works. Starting from the repair
trajectory $Traj_R$, \algo progressively searches for functionally
unnecessary edits by constructing and evaluating \emph{counterfactual
patches}. Rather than reasoning over individual modifications from the outset,
\algo first searches over larger removal candidates (e.g., an edit sequence like $\mathcal{E}$), allowing a
single successful validation to eliminate many edit actions simultaneously.
Only when no further removals are possible does the search refine to
smaller candidate units, thereby reducing the number of expensive
validation executions.

The search consists of three successive phases corresponding to the
hierarchy
$\textit{edit sequence} \rightarrow \textit{file} \rightarrow
\textit{edit action}$ (see the outer loop, Line~\ref{alg:outerloop}).
At the beginning of each phase, the surviving edits are \emph{regrouped}
(Line~\ref{alg:regroup}) according to the current search granularity (\textit{edit sequence},  \textit{file}, or \textit{edit action}),
thereby defining the ``unit" of a candidate removal for that phase.
During sequence-level search, each candidate consists of the edits
introduced between two consecutive feedback requests (See row 1 of~\Cref{fig:trim-example}).
During file-level search, the surviving edits are repartitioned so that
each candidate contains all edits affecting a particular file,
irrespective of when they were introduced (See row 2 of~\Cref{fig:trim-example}).
Finally, edit action-level search treats every surviving edit action as an independent
candidate, enabling fine-grained minimization (See row 3 of~\Cref{fig:trim-example}).
Thus, while the grouping of modifications changes across phases, the underlying
search procedure remains identical.

For each grouping, \algo performs the same iterative elimination
procedure. The algorithm traverses the candidate units in reverse
trajectory order and temporarily removes one unit at a time,
thereby constructing a \emph{counterfactual patch} that represents the ``hypothetical repair" had that candidate never been introduced.

\textit{Counterfactual Validation.} 
The counterfactual is accepted only if (i) it continues to satisfy the
task-specific test suite $TF$, thereby preserving the execution
invariant, and (ii) it produces a strictly smaller patch than the
current solution (Line~\ref{alg:accept}).
Otherwise, the candidate is rejected and the removed edits are restored. 
The elimination process repeats until no additional candidate units can
be removed at the current granularity (Line~\ref{alg:termination}),
thereby reaching a local fixpoint. The last row of~\Cref{fig:trim-example}) shows the final required edit actions following this algorithm. 

\subsubsection*{One-Minimality Guarantee}
\label{subsec:one_minimal}

The hierarchical search above efficiently removes unnecessary edits.
However, a single pass does not necessarily recover a
\emph{one-minimal} edit subset, since an edit that is initially
irremovable may become removable after other edits are eliminated. 
To guarantee one-minimality, \algo\ optionally enables the
\textit{oneMin} flag. When enabled, each search phase repeats until no
additional candidate can be removed (Line~\ref{alg:termination}).
Because every accepted removal strictly reduces the patch while
preserving the execution invariant, termination guarantees that no
remaining edit can be removed while still satisfying the task-specific
test suite $TF$. When \textit{oneMin} is disabled (\algo-NG), each
granularity is explored only once, reducing validation cost at the
expense of this guarantee.

\subsubsection*{Algorithmic Analysis}
\label{subsec:algorithmic_analysis}

The dominant cost of patch
minimization is executing the task-specific test suite $TF$. We therefore
measure the \textbf{cost complexity} $\mathcal{C}$ of \algo as the number of
$TF$ executions. 
 Without the one-minimality guarantee (\algoNG), each search phase performs a
single pass over its candidate units, yielding $
\mathcal{C}_{\algoNG}
=
\mathcal{O}(|Edit Seq|+|File|+|Edit Action|)
=
\mathcal{O}(|Edit Action|),
$ since $|Edit Action| \gg |Edit Seq| > |File|$.

With the one-minimality guarantee enabled (\algoG), each phase repeats until
reaching a local fixpoint. In the worst case, each candidate may be revisited
once for every remaining candidate, giving $
\mathcal{C}_{\algoG}
=
\mathcal{O}(|Edit Seq|^2+|File|^2+|Edit Action|^2)
=
\mathcal{O}(|Edit Action|^2).
$

Thus, enabling one-minimality increases the worst-case cost complexity from
linear to quadratic in the number of edits. As shown in
Section~\ref{sec:expr}, however, the practical overhead is much smaller,
making \algoG empirically close to linear while providing the stronger
guarantee. 
The same analysis and results apply to \textbf{runtime complexity} for both \algoG and \algoNG.

\section{Experimental Design}
\label{sec:expr}


\subsection{Study Subjects}
\label{subsec:rq_setup}

We evaluate \algo\ across two benchmarks and four agentic scaffolds to
assess its effectiveness, efficiency, and generality.

\textbf{Live-kBench}~\cite{huang2026outrunningllmcutoffslive} contains $534$
recent Linux kernel vulnerabilities for evaluating security-critical repair
in a large C codebase. We use it as our primary benchmark since Patch
Minimization is particularly important for security fixes, where smaller,
more focused patches simplify review and reduce unnecessary attack surface.
As a post-processing technique, \algo\ applies only to successful repairs.
Across scaffolds, this yields $433$ repaired bugs, of which $140$ are trivially one-minimal (single edit action), leaving $293$ bugs for evaluation.

\textbf{SWE-Bench-Verified}~\cite{jimenez2023swe} contains $500$ manually
verified GitHub issues from popular Python repositories. We evaluate
\algoG\ (edit-action level) on $333$ \swe (using Claude-Sonnet-4) repair trajectories
downloaded from the SWE-Bench leaderboard, all verified to pass the hidden
oracle.

We evaluate four state-of-the-art repair agents:
\crashfixer~\cite{mathai2025crashfixer}, \swe,
\mini~\cite{yang2024swe}, and
\openhands~\cite{wang2024openhands}. Across all
benchmark--scaffold combinations, \algo\ is evaluated on $4{,}544$ repair
trajectories generated using models including Gemini-3-Pro and
Claude-Opus-4.5.

\subsection{Evaluation Metrics}
\label{sec:eval_metric}


\textbf{(i) \aislop Reduction (\patchdiff).}
For each benchmark--scaffold combination, we quantify the amount of
\aislop\ removed by a minimization method (\algo or a baseline) as the
average normalized reduction in the patch length:

\begin{equation}
\Delta_{\text{Slop}} (\%)
=
\frac{
\operatorname{avg}\left(
  \operatorname{len}(\mathcal{AP})- \operatorname{len}(\mathcal{MP})
  \right)
}
{
\operatorname{avg}\!\left(\operatorname{len}(\mathcal{AP})\right)
} \times 100
\end{equation}

Here, $\mathcal{AP}$ denotes the original agent-generated patch and
$\mathcal{MP}$ the minimized patch. Higher \patchdiff\ values indicate that
more \aislop\ has been removed.

During minimization, \algo\ uses only the repair agent's test suite $TF$ as
its correctness signal. For Live-kBench, $TF$ is the crash-reproducing
script; for SWE-Bench-Verified, it consists of the tests generated and
executed by the agent during repair. Unless otherwise stated, all
\patchdiff\ results use only $TF$ for validation. The hidden oracle tests
described below are used solely for evaluation.

\textbf{(ii) Oracle Tests.}
Oracle tests are used only for evaluation and are never available to
\algo\ during minimization. For Live-kBench, the oracle combines the
crash-reproducing script with an LLM judge that determines semantic
equivalence to the developer patch. Following
\citet{huang2026outrunningllmcutoffslive}, we use Gemini-3-Flash with
nine independent judgments and majority voting. For
SWE-Bench-Verified, the oracle is the benchmark's hidden
\emph{fail-to-pass} test suite, which directly validates functional
correctness.

\textbf{(iii) Cost of Minimization.}
As discussed in Section~\ref{subsec:algorithmic_analysis}, 
we measure cost as the number of $TF$ executions, treating each
execution as one unit regardless of the number of test cases it contains.
Unlike wall-clock time, this machine-independent metric reflects the
algorithm's intrinsic efficiency and enables fair comparison across
execution environments. Within a benchmark, it is directly proportional to
both execution time and hardware cost.




\subsection{Baselines}
\label{subsec:baselines}

We compare \algo against two classes of baselines: \emph{agentic}
minimization and \emph{deterministic} patch minimization.

\textbf{Agentic Minimization.}
Our first baseline evaluates whether 
agents can minimize their own patches without an explicit minimization algorithm. We use \mini\ primarily with Gemini-3-Flash as the minimization agent, with a smaller focused evaluation using Sonnet 4.6. The agent is given the
same execution environment as \algo, including access to the task-specific
test suite $TF$ for validating candidates. Each minimization run is
allocated a budget of 6 hours and \$5.6 (following~\cite{huang2026outrunningllmcutoffslive}). We evaluate three prompting
configurations:
(a) \emph{Only Diff}, where the agent receives the final patch;
(b) \emph{Only Traj}, where it receives  the repair trajectory; and
(c) \emph{Both}, where it receives both the final patch and repair
trajectory.

\textbf{Deterministic Minimization.}
Our second baseline is a deterministic adaptation of Delta Debugging
(DD-Hunk)~\cite{10.1145/1134285.1134307}. Since traditional Delta Debugging
operates over program inputs rather than agent-generated patches, we modify
it to perform patch minimization over patch hunks. Like
\algo, DD-Hunk validates candidates patch using the task-specific test
suite $TF$, enabling a direct comparison between trajectory-guided and
direct patch-based minimization strategies.

\textbf{\algo Variants.}
We evaluate two variants of \algo. \algoNG performs a single pass at each
granularity, whereas \algoG repeats until no further reductions are possible,
guaranteeing a one-minimal solution. We also report results after
\emph{Sequence} (edit-sequence minimization), \emph{Edit} (full search), and
\emph{Hybrid}, which uses \emph{Sequence} when it yields a single edit
sequence and otherwise uses \emph{Edit}.


\section{Results}
\label{sec:result}

\subsection*{RQ1: How effectively does \algo\ reduce \aislop\ for security program repair?}
\phantomsection
\label{subsec:rq1}


\noindent\textbf{Approach.}
We evaluate both \algoG and \algoNG on the Live-kBench study subjects
described in Section~\ref{subsec:rq_setup}. We additionally report the
intermediate results after each stage of the hierarchical search
(\emph{Sequence}, \emph{Hybrid}, and \emph{Edit}) to quantify the
contribution of each search granularity. We measure effectiveness using
\patchdiff, oracle correctness (See~\Cref{sec:eval_metric}), and efficiency (\ie cost of minimization) 
using the number of test-suite executions ($TF$).

\begin{table}[h]
\centering
\caption{Live-kBench \% \patchdiff using (i) \algoG with different minimization granularity \& (ii) Agentic Minimization (+ Filters) with different minimization settings. Claude Sonnet $4.6$ run for one scaffold due to resource constraints.
}
\label{tab:agent_min_results}
\resizebox{0.9\linewidth}{!}{
\begin{tabular}{lcccc}
\toprule
\rowcolor{gray!15}
 & \textbf{Crash-} & \textbf{SWE } & \textbf{Open-} & \textbf{Mini-SWE } \\
\rowcolor{gray!15}
 & \textbf{Fixer} & \textbf{Agent} & \textbf{Hands} & \textbf{ Agent} \\
\midrule
\rowcolor{gray!15}
\midrule
\multicolumn{3}{l}{ \textbf{\algoG Minimization \patchdiff}\% } & & \\
Sequence  & 16.4\%  & 5.6\%  & 13.3\%   & 17.4\%         \\
\textit{Hybrid} & 24.7\%    & 8.9\%     & 19\%    & 18\%  \\
Full  & \textbf{32.9}\%    & \textbf{17.9\%}    & \textbf{26.6\%}    & \textbf{26\%}  \\
$\times$ Ratio $\uparrow$ & \textbf{3.1x}       & \textbf{1.6x}      & \textbf{1.9x}      & \textbf{2.0x} \\
\midrule
\rowcolor{gray!15}
\midrule
\multicolumn{3}{l}{ \textbf{Agentic Minimization \patchdiff\% (Gemini 3 Flash)} } & & \\
Only Diff & 5.5\% & 2.1\% & 3.4\% & 2.4\% \\
Only Traj & \textbf{10.5\%} & \textbf{11.3\%} & \textbf{13.4\%} & \textbf{12.6\%} \\
Both & 2.8\% & 2.4\% & 2.4\% & 6.4\% \\
\midrule
\rowcolor{gray!15}
\multicolumn{4}{l}{ \textbf{Agentic Minimization \patchdiff\% (Claude Sonnet 4.6)} } & \\
Only Diff & 8.9\% & - & - & - \\
Only Traj & \textbf{9.9\%} & - & - & - \\
Both & 9.6\% & - & - & - \\
\bottomrule
\end{tabular}
}
\end{table}

\noindent\textbf{Results.}

\noindent\textbf{\aislop\ Reduction ($\Delta_{\text{Slop}}$\%).}
Table~\ref{tab:agent_min_results} shows that the full edit-level version of
\algoG consistently achieves the largest reduction in \aislop, removing
$17.9$--$32.9\%$ of the agent-generated patch across all four repair agents.
The results also validate the hierarchical design: sequence-level
minimization removes the least \aislop, edit-level the most, with the hybrid
variant consistently lying between the two.

\noindent\textbf{Oracle Performance.}
Despite its aggressive minimization, \algoG preserves oracle correctness
(Table~\ref{tab:oracle_check}). It even improves \crashfixer\ by
$1.0$--$2.6\%$, producing patches that more closely resemble the developer
solution. For the remaining agents, oracle performance remains essentially
unchanged, with at most a $\sim$1\% drop for \openhands\ under edit-level
minimization. Since our evaluation additionally requires every minimized
patch to satisfy the hidden oracle, these results represent a conservative
lower bound on \algo's effectiveness.

\begin{table}[h]
\centering
\caption{Live-kBench Oracle Performance for \algoG: \% of patches equivalent to human fix \& that resolve the crash 
}
\label{tab:oracle_check}
\resizebox{0.8\linewidth}{!}{
\begin{tabular}{lcccc}
\toprule
\rowcolor{gray!15}
 & \multicolumn{4}{c}{\textbf{Equivalent Patch \& Crash Resolved Rate}} \\
\cmidrule(lr){2-5}
\rowcolor{gray!15}
\textbf{Strategy/} & \textbf{Crash}& 
\textbf{SWE} & \textbf{Open} & \textbf{MiniSWE} \\
\rowcolor{gray!15}
\textbf{Phase} & \textbf{Fixer} & 
\textbf{Agent} &  \textbf{Hands} &  \textbf{Agent} \\
\midrule
\agentp & 22.47\% & 23.03\% & 21.54\% & 20.04\% \\
\midrule
Sequence  & 24.91\%  & 23.41\%  & 21.72\%  & 20.22\%  \\
          & \textcolor{improvgreen}{(+2.44)} & \textcolor{improvgreen}{(+0.38)} & \textcolor{improvgreen}{(+0.18)} & \textcolor{improvgreen}{(+0.18)} \\
\midrule
\textit{Hybrid}   & 25.09\%  & 23.41\%  & 21.72\%  & 20.41\%  \\
    & \textcolor{improvgreen}{(+2.62)} & \textcolor{improvgreen}{(+0.38)} & \textcolor{improvgreen}{(+0.18)} &  \textcolor{improvgreen}{(+0.37)} \\
\midrule
Full  & 23.41\%  & 22.85\%  & 20.41\%  & 19.66\%  \\
      &  \textcolor{improvgreen}{(+0.94)} &   \textcolor{red}{(-0.18)} &\textcolor{red}{(-1.13)} & \textcolor{red}{(-0.38)} \\

\bottomrule
\end{tabular}
}
\vspace{-6mm}
\end{table}

\noindent\textbf{\algoG vs.~\algoNG.}
Figure~\ref{fig:cost} shows that relaxing the one-minimality guarantee has
little practical impact. \algoNG achieves nearly identical \patchdiff\
($32.0\%$ vs.~$32.9\%$), average edits per patch ($1.35$ for both), and
oracle correctness, producing identical minimized patches for $96.4\%$ of
repair trajectories while reducing validation cost by $\sim$8\%
($2.4$k vs.~$2.6$k kernel jobs). Thus, \algoNG\ offers a favorable
cost--quality trade-off, whereas \algoG\ remains preferable when a
one-minimal guarantee is required.

Figure~\ref{fig:cost} also illustrates the broader cost--quality trade-off.
Sequence-level minimization provides the lowest cost ($\sim$0.9k jobs) with $11.3\%$ \patchdiff, while edit-level minimization achieves the
largest \patchdiff ($32.9\%$) at $\sim$2.5k jobs, allowing users to select appropriately for their validation budget.

\begin{rqsummary}{RQ1 Summary}
\small{
\textbf{\algo\ effectively and efficiently removes \aislop.} 
\algo\ removes up to 32.9\% of unnecessary edits with negligible
oracle loss. \algoNG $\approx$ \algoG in minimization quality while reducing
validation cost and giving a linear time guarantee.
}
\end{rqsummary}

\subsection*{RQ2: How does \algo\ compare against agentic minimization in reducing \aislop?}
\phantomsection
\label{subsec:rq2}

\begin{table}[h]
\centering
\caption{\% of Agentic Minimized Patches (Gemini-3-Flash) in Live-kBench filtered out due to failing criteria. Crash (\crashfixer), Mini (\mini), SWE (\swe), and OH (\openhands) are the four scaffolds \& total number of trajectories is $4.5k$.
}
\label{tab:filtered_out}
\resizebox{\linewidth}{!}{
\begin{tabular}{llccccc}
\toprule
\rowcolor{gray!15}
 & & \multicolumn{4}{c}{\textbf{\% of Invalid Minimizations in each Category}} & \\
\cmidrule(lr){3-6}
\rowcolor{gray!15}
\textbf{\makecell{Agentic \\ Setting}} & \textbf{Scaffold} & \textbf{Anomaly} & \textbf{\makecell{File \\ Mismatch}} & \textbf{\makecell{Bug \\ Triggered}} & \textbf{\makecell{Patch \\ Length $\uparrow$}} & \textbf{Total} \\
\midrule
\multirow{4}{*}{\makecell{Only \\ Diff}} & Crash & 2.1\% & 0.5\%  & 2.5\%  & 0.5\%  & 5.6\%  \\
 & SWE & 1.0\% & 1.3\%  & 3.8\%  & 0.7\%  & \textbf{6.8\%}  \\
 & Mini  & 1.0\% & 0.5\%  & 1.3\%  & 1.0\%  & 3.8\%  \\
 & OH  & 1.8\% & 0.2\%  & 3.0\%  & 1.2\%  & 6.2\%  \\
\midrule
\multirow{4}{*}{\makecell{Only \\ Traj}} & Crash & 2.1\% & 8.0\%  & 11.1\% & 11.9\% & 33.1\% \\
 & SWE & 3.2\% & 14.6\% & 14.6\% & 12.5\% & \textbf{44.9\%} \\
 & Mini  & 1.8\% & 4.8\%  & 8.3\%  & 10.6\% & 25.5\% \\
 & OH  & 3.0\% & 5.5\%  & 12.2\% & 16.5\% & 37.2\% \\
\midrule
\multirow{4}{*}{\makecell{Both}} & Crash & 3.0\% & 4.8\%  & 4.1\%  & 2.3\%  & 14.2\% \\
 & SWE & 1.6\% & 6.2\%  & 6.1\%  & 3.6\%  & \textbf{17.5\%} \\
 & Mini  & 1.5\% & 3.0\%  & 3.8\%  & 3.8\%  & 12.1\% \\
 & OH  & 1.2\% & 3.1\%  & 5.2\%  & 2.1\%  & 11.6\% \\
\bottomrule
\end{tabular}
}
\end{table}


\noindent\textbf{Approach.}
We compare \algo against the three agentic minimization baselines described
in Section~\ref{subsec:baselines}. All methods are evaluated using the same
validation environment and execution budget, and are compared using
\patchdiff.

\noindent\textbf{Results.}
Table~\ref{tab:agent_min_results} shows that \algo\
substantially outperforms agentic minimization across all four scaffolds. The edit-level version of \algoG achieves
\patchdiff of $17.9$--$32.9\%$, corresponding to a
$1.6\times$--$3.1\times$ improvement over the best-performing
agentic baseline. While prompting an LLM with the final patch
(\textit{Only Diff}), the repair trajectory (\textit{Only Traj}),
or both occasionally removes some \aislop, none consistently
approaches the minimization achieved by \algo.

Unlike \algo, which deterministically returns a valid minimized patch,
agentic minimization fails in $3.8$--$44.9\%$ of cases
(Table~\ref{tab:filtered_out}), including patch inflation, file
mismatches, and bug reintroduction. We conservatively count these
failures as no minimization when computing
Table~\ref{tab:agent_min_results}; even under this favorable fallback,
the agentic baselines are substantially worse.

These results suggest that Patch Minimization is fundamentally a
\emph{search} problem rather than a \emph{generation} problem:
trajectory-guided counterfactual search is both more effective and reliable than prompting an LLM to rewrite its own patch.

\begin{rqsummary}{RQ2 Summary}
\small{
\textbf{\algo\ outperforms agentic minimization.} It removes
$1.6\times$--$3.1\times$ more \aislop\ while producing more reliable patches,
showing that structured counterfactual search is more effective and reliable
than agentic patch revision.}
\end{rqsummary}

\subsection*{RQ3: How does \algo\ compare against deterministic patch minimization algorithms? }
\phantomsection
\label{subsec:rq3}


\noindent\textbf{Approach.}
We compare \algo against the deterministic DD-Hunk baseline introduced in
Section~\ref{subsec:baselines}. We evaluate minimization quality
(\patchdiff) and computational cost ($TF$ executions) to understand the
benefits of trajectory-guided search.

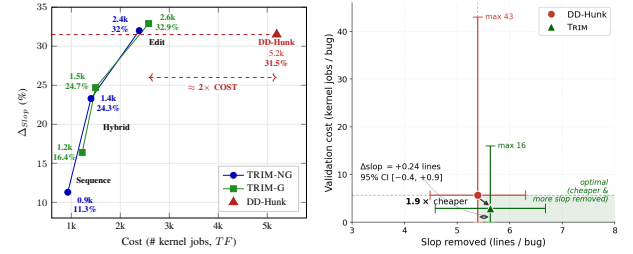
\begin{figure}[h]
\centering
\resizebox{0.45\columnwidth}{!}{
\begin{tikzpicture}
\begin{axis}[
    width=\columnwidth, height=7.6cm,
    xlabel={Cost (\# kernel jobs, $TF$)},
    ylabel={\patchdiff\ (\%)},
    xmin=600, xmax=5800,
    ymin=8,  ymax=35,
    xtick={1000,2000,3000,4000,5000},
    xticklabels={1k,2k,3k,4k,5k},
    grid=both,
    grid style={gray!18},
    legend pos=south east,
    legend cell align=left,
    legend style={font=\footnotesize, fill=white, draw=gray!40},
    label style={font=\small},
    tick label style={font=\footnotesize},
]
\addplot[mark=*, thick, blue!70!black, mark size=2.4pt, line join=round] coordinates {
    (927, 11.3) (1401, 23.3) (2386, 32)
};
\addlegendentry{\algoNG}
\addplot[mark=square*, thick, improvgreen, mark size=2.4pt] coordinates {
    (1221, 16.4) (1495, 24.7) (2578, 32.9)
};
\addlegendentry{\algoG}
\addplot[only marks, mark=triangle*, regressred, mark size=4pt] coordinates {
    (5192, 31.5)
};
\addlegendentry{DD-Hunk}
\draw[regressred, dashed, semithick] (axis cs:600,31.5) -- (axis cs:5192,31.5);
\node[font=\scriptsize\bfseries, anchor=west] at (axis cs:1015,12.8) {Sequence};
\node[font=\scriptsize\bfseries, anchor=west] at (axis cs:1560,19.6) {Hybrid};
\node[font=\scriptsize\bfseries, anchor=west] at (axis cs:2500,30.5) {Edit};
\node[font=\scriptsize\bfseries, blue!70!black, align=center, anchor=north west] at (axis cs:955,11.3)  {0.9k\\11.3\%};
\node[font=\scriptsize\bfseries, blue!70!black, align=center, anchor=north west] at (axis cs:1430,24.3) {1.4k\\24.3\%};
\node[font=\scriptsize\bfseries, blue!70!black, align=center, anchor=north west, fill=white, inner sep=0.8pt] at (axis cs:1800,34) {2.4k\\32\%};
\node[font=\scriptsize\bfseries, improvgreen, align=center, anchor=east]       at (axis cs:1200,16.4) {1.2k\\16.4\%};
\node[font=\scriptsize\bfseries, improvgreen, align=center, anchor=south east] at (axis cs:1455,24) {1.5k\\24.7\%};
\node[font=\scriptsize\bfseries, improvgreen, align=center, anchor=south east] at (axis cs:3300,31.5) {2.6k\\32.9\%};
\node[font=\scriptsize\bfseries, regressred, align=center, anchor=north] at (axis cs:5192,31.5) {DD-Hunk\\[1pt]\mdseries 5.2k\\31.5\%};
\draw[<->, regressred, dashed, thick]
    (axis cs:2578,26.0) -- (axis cs:5192,26.0);
\node[regressred, font=\scriptsize\bfseries, fill=white, inner sep=1.5pt, anchor=south]
    at (axis cs:3885,24) {$\approx$ 2$\times$ COST};
\end{axis}
\end{tikzpicture}

}
\includegraphics[width=0.45\columnwidth]{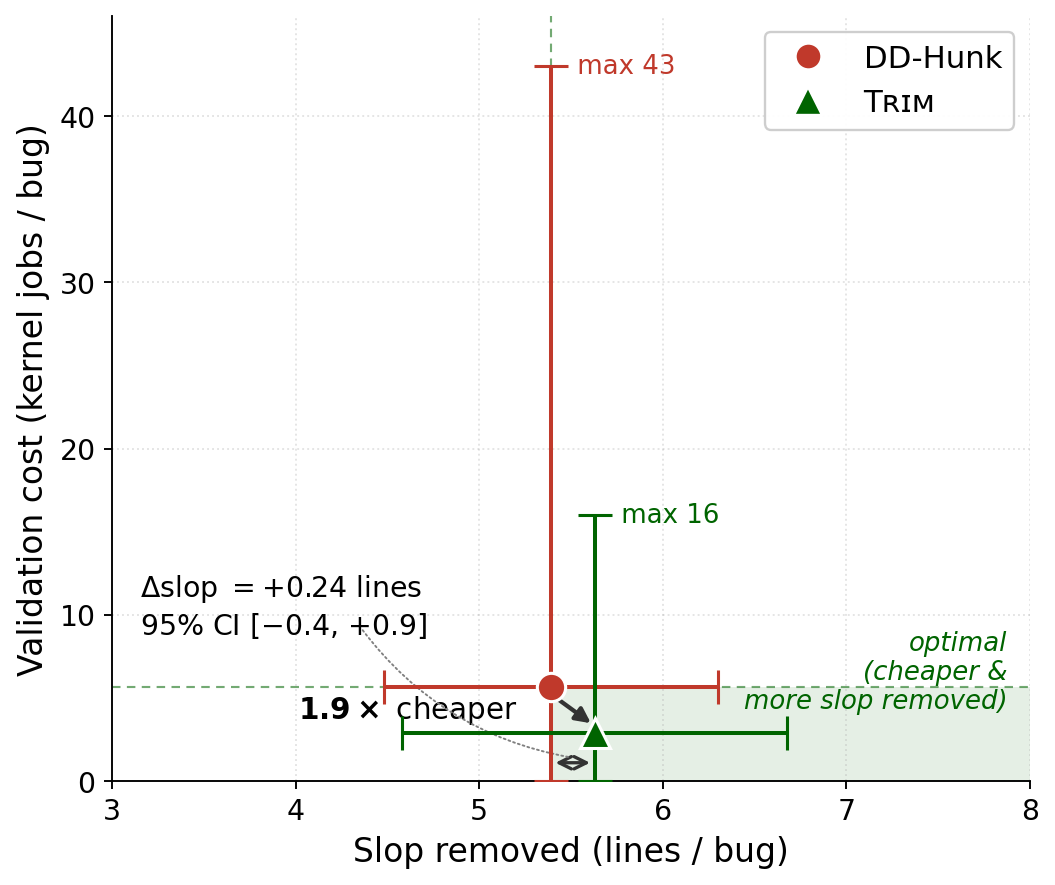}
\caption{\small{\textbf{Left.}~\patchdiff\% vs Cost for \algo's variants and DD-Hunk on \crashfixer\ trajectories (Live-kBench). 
\algo variants reach comparable \patchdiff at half the cost of DD-Hunk.
~\textbf{Right.}~\algo\ vs.\ DD-Hunks. \algo\ removes similar slop ($\Delta$=+0.24 lines, 95\% CI [-0.4,+0.9]) at $1.9\times$ lower cost. The shaded green region indicates the better setting: more slop removed at cheaper cost.}}
\label{fig:cost}
\vspace{-3mm}
\end{figure}

\noindent\textbf{Results.}
Figure~\ref{fig:cost} (left) shows that \algoG\ (edit level) achieves
essentially the same minimization quality as DD-Hunk, removing $32.9\%$
versus $31.5\%$ of \aislop\ ($5.63$ vs.\ $5.39$ lines per bug;
$p=0.50$). However, \algo\ requires only $\sim$2.6k kernel validations,
compared to DD-Hunk's $\sim$5.2k---a $1.9\times$ reduction in validation
cost. Figure~\ref{fig:cost} (right) further shows that \algo\ is substantially
more predictable. DD-Hunk requires up to $43$ validations per bug, whereas
\algo\ never exceeds $16$. Since each kernel validation takes
$\sim$30 minutes, bounding the number of validations significantly reduces
worst-case execution time and compute cost.

These results demonstrate the benefit of trajectory-guided hierarchical
search: by eliminating large groups of related edits before refining
individual edits, \algo\ matches exhaustive hunk-level minimization with nearly half the validation budget.



\begin{rqsummary}{RQ3 Summary}
\small{
\textbf{\algo's hierarchical search matches delta-debugging-based hunk minimization at much
lower cost.} Compared to DD-Hunk, \algo\ achieves statistically equivalent
\aislop\ reduction while requiring \textbf{1.9$\times$ fewer} validation
executions and substantially lower worst-case cost.
}
\end{rqsummary}


\subsection*{RQ4: Does \algo\ generalize to general-purpose program repair benchmarks and other ``static" \aislop metrics?}
\phantomsection
\label{subsec:rq4}


\noindent\textbf{Approach.}
We evaluate \algoG\ (edit-action level) on $333$ \swe repair trajectories (on Claude-Sonnet-4) from SWE-Bench-Verified. Unlike Live-kBench, these trajectories
contain heterogeneous feedback requests and evolving tests. We 
adapt only trajectory preprocessing to construct the validation signal
($TF$), leaving the core minimization algorithm unchanged. We then measure
\patchdiff\ using $TF$ and verify correctness against the hidden oracle. Additionally, we also check if other ``static" definitions of \aislop like \textbf{verbosity} \cite{orlanski2026slopcodebench} reduce when running \algo.

\noindent\textbf{Results.}
Table~\ref{tab:min_res} reports \patchdiff\ after removing changes
introduced by the SWE-Bench evaluation infrastructure. The conservative
measure, restricted to pre-existing repository files, yields a
$20.0\%$ \patchdiff\ ($23.6\%$ of edit actions and $17.8\%$ of hunks
removed), comparable to the $17.8\%$ achieved on Live-kBench for the same
\swe\ scaffold (Table~\ref{tab:agent_min_results}). Including all modified
files gives an upper bound of $63.5\%$.

Importantly, \algo\ preserves oracle correctness for $327/330$ ($99.1\%$)
successful repairs, with only three regressions, and transforms $18$ patches
into the developer-written solution. These results demonstrate that \algo\
generalizes from Linux kernel security repair to repository-level program
repair without modifying its core hierarchical minimization algorithm.

Additionally, \algo-generated minimized patches also substantially reduce other static measures of \aislop, like verbosity \cite{orlanski2026slop}, as shown in Table \ref{tab:verbo}.

\begin{table*}[t]
  \centering
  \caption{Effect of \algo\ on SWE-Bench-Verified ($327$ oracle-preserved patches).}
  \label{tab:swebench_results}

  \begin{minipage}[t]{0.32\textwidth}
  \centering
  \setcounter{subtable}{0}
  \begin{subtable}[t]{\linewidth}
  \centering
  \caption{Overall minimization.}
  \label{tab:min_res}
  \resizebox{\linewidth}{!}{
  \begin{tabular}{lcccc}
  \toprule
  \rowcolor{gray!15}
  \textbf{Metric} &
  \textbf{Edits} &
  \textbf{Hunks} &
  \textbf{\makecell{Lines\\(Total)}} &
  \textbf{\makecell{Lines\\(Mod.\ files)}} \\
  \midrule
  \patchdiff \% & 23.6\% & 17.8\% & \textbf{63.5\%} & 20.0\% \\
  \bottomrule
  \end{tabular}
  }
  \end{subtable}

  \vspace{2mm}

  \setcounter{subtable}{3}
  \begin{subtable}[t]{\linewidth}
  \centering
  \caption{\% of \agentp-introduced verbosity \cite{orlanski2026slopcodebench} removed, across the 90 minimized instances.}
  \label{tab:verbo}
  \resizebox{0.8\linewidth}{!}{
  \begin{tabular}{lccc}
  \toprule
  \rowcolor{gray!15}
  \textbf{File type} &
  \textbf{\makecell{Intro.\\Verb}} &
  \textbf{\makecell{Rem.\\Verb}} &
  \textbf{\makecell{\%\\Removed}} \\
  \midrule
  Scratch repro & 1006 & 1006 & 100\% \\
  Source        & 650  & 122  & 19\%  \\
  Tests         & 238  & 121  & 51\%  \\
  \midrule
  \textbf{Total} & \textbf{1894} & \textbf{1249} & \textbf{66\%} \\
  \bottomrule
  \end{tabular}
  }
  \end{subtable}
  \end{minipage}
  \hfill
  \begin{minipage}[t]{0.33\textwidth}
  \centering
  \setcounter{subtable}{1}
  \begin{subtable}[t]{\linewidth}
  \centering
  \caption{Breakdown by file type.}
  \resizebox{\linewidth}{!}{
  \begin{tabular}{lcc}
  \toprule
  \rowcolor{gray!15}
  \textbf{File type} &
  \textbf{\patchdiff\%} &
  \textbf{\makecell{Share of\\removed lines}} \\
  \midrule
  Scratch repro scripts & 100\% & 49.6\% \\
  Build / config        & 100\% & 35.8\% \\
  Documentation         & 100\% & 0.2\% \\
  Source                & 20.4\% & 11.5\% \\
  Tests                 & 18.6\% & 2.9\% \\
  \bottomrule
  \end{tabular}
  }
  \end{subtable}
  \end{minipage}
  \hfill
  \begin{minipage}[t]{0.31\textwidth}
  \centering
  \setcounter{subtable}{2}
  \begin{subtable}[t]{\linewidth}
  \centering
  \caption{Structural changes.}
  \resizebox{\linewidth}{!}{
  \begin{tabular}{lcc}
  \toprule
  \rowcolor{gray!15}
  \textbf{Outcome} & \textbf{\# Cases} & \textbf{\# Files} \\
  \midrule
  Unchanged ($\mathcal{MP}{=}\mathcal{AP}$) & 237 & --- \\
  Minimized                     & 90  &     \\
  \quad Single file, trimmed    & 54  & --- \\
  \quad Multi-file, same count  & 12  & --- \\
  \quad Multi-file, fewer files & 24  & 62  \\
  \qquad Scratch                &     & 32  \\
  \qquad Source                 &     & 14  \\
  \qquad Tests                  &     & 7   \\
  \qquad Build/config           &     & 6   \\
  \qquad Docs                   &     & 3   \\
  \midrule
  \textbf{Total}                & \textbf{327} & \textbf{62} \\
  \bottomrule
  \end{tabular}
  }
  \end{subtable}
  \end{minipage}

  \vspace{-3mm}
  \end{table*}
\begin{rqsummary}{RQ4 Summary}
\small{
\textbf{\algo\ generalizes across domains,} achieving comparable \aislop\ reduction on
SWE-Bench-Verified while preserving oracle correctness on
\textbf{99.1\%} successful repairs. \algo also reduces other static definitions of \aislop, like verbosity.
}
\end{rqsummary}

\section{Related Work}




\textbf{AI Slop.} In NLP, the term ``slop" characterizes low-quality, verbose, or stylistically degraded AI-generated \emph{text}~\cite{shaib2026measuringaisloptext}.
More relevant to us, a growing line of SE work studies the
\emph{quality degradation} of agent-generated code --- redundant
constructions, duplication, and the concentration of complexity in
already-complex functions --- and shows that these properties worsen as
agents extend their own code over long
horizons~\cite{orlanski2026slop,dou2026wrong,abbassi2025taxonomy}. 
That work terms slop as a \emph{static quality property} of the code an agent retains: verbose or over-complex code that nonetheless still performs work, and that is measured directly on the surviving source. 
Our notion is orthogonal. \textsc{CodeSlop} is \emph{functional redundancy} --- edits that can be removed \emph{in their entirety} while the patch still passes its tests --- and it is a \emph{structural artifact of the agent's search trajectory} rather than a stylistic property of any one line. 
The two genuinely differ: a removable edit may be perfectly clean code (so static
quality metrics wouldn't flag it), while an eroded but irreducible
function is not removable at all. 
To our knowledge, we are the first to (i) formalize slop as removable functional redundancy within an agent patch, and (ii) propose algorithms that \emph{remove} it while preserving test-defined correctness. 

\textbf{Minimization.}  Reducing an artifact to a minimal form, such that it still satisfies a property, has been explored in research like  Delta Debugging \cite{10.1145/318774.318946}---which isolates a $1$-minimal failure-inducing input, and its hierarchical variant HDD~\cite{10.1145/1134285.1134307}, which exploits syntactic structure for efficiency. 
These works caused a flurry of \textit{program-reducing} research, like test-case reducers such as C-Reduce~\cite{10.1145/2345156.2254104} and Perses~\cite{10.1145/3180155.3180236}, \textit{program-debloating}  works like Chisel~\cite{10.1145/3243734.3243838}, and other related debloating works minimizing at the container~\cite{10.1145/3106237.3106271}, bytecode~\cite{inproceedings}, or
load-time~\cite{217642} granularity.
A recent thread ~\cite{jia2026compressingcodecontextllmbased} minimizes the \emph{input context} to a coding agent: 
extracting a minimal yet sufficient context that allows a model to synthesize a correct patch.
In contrast, \algo targets a different objective. Rather than minimizing source code or LLM context, \algo minimizes the \emph{patch} an agent generates, removing redundant edits that agents accumulate while resolving an issue. 


\textbf{Agents.} Agents that condense all the relevant context into a single prompt tend to perform poorly on repository-level benchmarks \citep{mathai2024kgym, jimenez2023swe}. 
This limitation motivated a shift toward autonomous SE agents: \swe \cite{yang2024swe} established the paradigm by granting LLMs access to computer-use tools, and its bash-only variant (\mini) now serves as a standard scaffold for evaluating LLMs on SWE-bench \citep{jimenez2023swe, swebenchleaderboards, yang2024swe}. 
A wave of subsequent agentic frameworks followed, among them \openhands \cite{wang2024openhands}, TRAE \cite{traeresearchteam2025traeagent}, {Live-SWE-agent}\xspace \cite{xia2025live}, CodeResearcher \cite{singh2025code}, and \crashfixer \cite{mathai2025crashfixer}. In this work, we analyze and minimize trajectories for four of the above coding agents and show that \algo can significantly reduce \aislop in their submitted patches.
\section{Threats to Validity}
\label{sec:discussion}
\label{sec:threats}

\subsection{Discussion}

Our results suggest that Patch Minimization is fundamentally a
\emph{search} problem rather than a \emph{generation} problem.
Trajectory-guided counterfactual search consistently outperforms prompting an
LLM to rewrite its own patch, while avoiding the reliability issues of
agentic minimization. Unlike Delta Debugging, which treats patches as
largely independent hunks, \algo\ exploits the dependencies naturally encoded
in repair trajectories, matching exhaustive hunk-level minimization at
roughly half the validation cost. Moreover, because \algo\ progressively
refines the search from coarse to fine granularity while maintaining a valid
patch throughout, it exposes a practical cost--quality trade-off, allowing
users to stop at any stage depending on their available validation budget.


\subsubsection{Internal Validity}

\textbf{Edit granularity.}
\algo's smallest removable unit is an atomic edit action. Thus, redundancy
within a single edit action cannot be eliminated, making the reported
\patchdiff\ a conservative lower bound. 

\textbf{Oracle imperfection.}
\algo\ relies on the task-specific test suite ($TF$) as a proxy for program
correctness, so untested regressions may go undetected. Empirically, oracle
regressions are rare (3/330 SWE-Bench patches and at most $\sim$1\% on
Live-kBench). In addition, Live-kBench semantic equivalence is evaluated
using an LLM judge, whose inherent bias may slightly affect the reported
absolute gains.

\subsubsection{Construct Validity}


\textbf{Conditioning and evaluation.}
\algo\ operates only on successful repairs, and our evaluation additionally
requires every minimized patch to satisfy a hidden oracle, although
deployment relies only on $TF$. Both choices make the reported effectiveness
a conservative lower bound.

\subsubsection{External Validity}

\textbf{Benchmark generality.}
Our evaluation spans both security-critical kernel repair and repository-level repair. While only trajectory preprocessing is benchmark-specific, other domains may exhibit different trajectory structures or testing workflows that affect \patchdiff performance.

\section{Conclusion}

In this work, we formally define the term \aislop, and equate the task of minimizing \aislop to \patchmin. We highlight that redundant edits in agent trajectories are one of the main root causes of the \aislop observed in agent patches. We then present \algo---a trajectory-aware minimization algorithm that minimizes edit-actions in agent trajectories. Empirically, we show that \algo uses this indirect technique to efficiently minimize up to $32.9\%$ of \aislop for half the cost of Delta Debugging.

\bibliographystyle{IEEEtranN}
\bibliography{main}

\end{document}